# Nutrient Supply to Planetary Biospheres from Anoxic Weathering of Mafic Oceanic Crust

**D. D. Syverson[1*], C. T. Reinhard[2,3,4], T. T. Isson[1,5], C. H. Holstege[1], J. A. R. Katchinoff[1], B. M. Tutolo[6], B. Etschmann[7], J. Brugger[7], and N. J. Planavsky[1,3]**

[1]Yale University, Department of Earth and Planetary Sciences, New Haven, CT, USA

[2]Georgia Institute of Technology, Earth and Atmospheric Sciences, Atlanta, GA, USA

[3]NASA Interdisciplinary Consortia for Astrobiology Research (ICAR), Alternative Earths Team

[4]NASA Nexus for Exoplanet System Science (NExSS)

[5]University of Waikato (Tauranga), Faculty of Science & Engineering, NZ

[6]University of Calgary, Department of Geosciences, Calgary, AB, CA

[7]Monash University, Earth, Atmosphere & Environment, Melbourne, VIC, AU

[*]Corresponding author: Drew Syverson (drew.syverson@yale.edu)

**Key Points:**

- Continental weathering is conventionally considered to be the only phosphorus source to planetary biospheres
- We experimentally demonstrate that weathering of ocean crust under anoxic conditions releases significant amounts of bioavailable phosphorus
- Habitable planets (including the earliest Earth) without subaerial continents may support significant biogenic gas fluxes to the atmosphere

## Abstract

Phosphorus is an essential element for life, and the phosphorous cycle is widely believed to be a key factor limiting the extent of Earth's biosphere and its impact on remotely detectable features of Earth's atmospheric chemistry. Continental weathering is conventionally considered to be the only source of bioavailable phosphorus to the marine biosphere, with submarine hydrothermal processes acting as a phosphorus sink. Here, we use a novel $^{29}Si$ tracer technique to demonstrate that alteration of submarine basalt under anoxic conditions leads to significant soluble phosphorus release, with an estimated ratio between phosphorus release and $CO_2$ consumption ($\sum PO_4^{3-}/\sum CO_2$) of 3.99 ± 1.03 µmol $mmol^{-1}$. This ratio is comparable to that of modern rivers, suggesting that submarine weathering under anoxic conditions is potentially a significant source of bioavailable phosphorus to planetary oceans and that volatile-rich Earth-like planets lacking exposed continents could develop robust biospheres capable of sustaining remotely detectable atmospheric biosignatures.

## Plain Language Summary

It is conventionally thought that continents above sea level are required in order for habitable planets to support a robust biosphere. We use experimental geochemistry and a simple model of biological cycling to show that this is incorrect, significantly expanding the possible range of planets that may host surface biospheres that would be detectable through telescope observations.

## 1. Introduction

Phosphorus (P) is a critical component of the genetic and energetic machinery of all life and plays key structural roles in most organisms. Indeed, recent debate has bolstered the case that P is not only essential for life on Earth but is also likely central for recognizable biochemistry more broadly (Erb et al., 2012; Reaves et al., 2012). Recent biogeochemical modeling and reconstructions of the evolution of marine phosphate ($PO_4^{3-}$) concentrations from Earth's rock record indicate that P has been the ultimate limiting nutrient for the biosphere throughout Earth's history (Derry, 2015; Laakso & Schrag, 2014, 2018), and it has also been argued that P would be expected to limit the extent of life on exoplanets where oxygenic photosynthesis has evolved (Reinhard et al., 2017). Therefore, a mechanistic understanding of how the global P cycle has changed through time is crucial both for a basic understanding of the history of life on our planet and in the development of predictive frameworks for the production and maintenance of exoplanet biosignatures (Meadows et al., 2018).

Marine $PO_4^{3-}$ concentrations are controlled through time by the interplay between the magnitude of P source(s) into the ocean and the efficiency of P burial. Continental weathering, an important long-term $CO_2$ sink that acts to regulate planetary climate, is also the only significant source of P to the modern oceans (Ruttenberg, 2003). Submarine weathering of basaltic oceanic crust, while also serving as a long-term $CO_2$ sink, currently acts a significant removal process of P from marine systems (L.A. Coogan & Gillis, 2013; McManus et al., 2019; Wheat et al., 2017; Wheat et al., 2003). For example, roughly 20% of the P sourced to the oceans today is removed through basalt weathering (Ruttenberg, 2003; Wheat et al., 2003). The remainder is removed in association with the burial of authigenic apatite and carbonate minerals, organic P, and iron oxides in marine sediments. It has also been proposed that anoxic and iron-rich (ferruginous) oceans, which were widespread on the early Earth (Poulton & Canfield, 2011; Song et al., 2017), lead to enhanced P scavenging through adsorption onto iron oxide minerals formed near the oxygenated ocean-atmosphere interface, through the precipitation of reduced iron-phosphate minerals, such as vivianite, directly from seawater, or through scavenging onto a range of other reduced Fe-bearing mineral phases (Bjerrum & Canfield, 2002; Derry, 2015; Johnson et al., 2020; Reinhard et al., 2017).

Phosphorus removal during basalt weathering may have been enhanced in Earth's past, given that it is likely that oceanic crust weathering played a more significant role in weathering and $CO_2$ sequestration prior to the emergence of continents above sea level and the proliferation of land plants in terrestrial ecosystems (Krissansen-Totton et al., 2018; Mills et al., 2014). However, previous work has neglected the possibility that dissolved P may become liberated into seawater during marine weathering of oceanic crust in the absence of dissolved $O_2$—as a natural result of limited $Fe^{2+}$ oxidation and subsequent P scavenging. Although mid-ocean ridge basalts (MORB) typically do not contain igneous apatite, $P^{5+}$ substitutes for $Si^{4+}$ in primary silicate minerals (Koritnig, 1965; Watson, 1980), and this P could potentially be released during submarine basalt weathering. If operative, this process would reshape our view of the evolution of the P cycle on Earth (Mills et al., 2014; Reinhard et al., 2017) and would become an important component of attempts to predict planetary P cycling on habitable exoplanets. Here, we provide direct experimental support for the idea that basalt weathering under anoxic marine conditions can be a significant source of bioavailable P to aqueous systems, compare our observations with modern seafloor weathering systems, and provide initial estimates of the potential global biogeochemical impacts of this process.

## 2. Materials and Methods

We use an enriched dissolved $^{29}SiO_2$ tracer to directly and accurately correlate the extent of primary silicate mineral dissolution with the amount of $PO_4^{3-}$ mobilized into seawater upon reaction with natural submarine basalt. We focus here on two sets of experiments, one utilizing fresh submarine basalt and one in which the fresh submarine basalt was treated with a reductive dissolution procedure designed to remove pre-existing $Fe^{3+}$-oxides associated with the partial oxidation during recovery from the seafloor (Mehra & Jackson, 1958). We combine extensive mineral characterization of the reactant basalt with time-series changes in solution chemistry and thermodynamic modeling to evaluate the efficacy of basalt weathering as a potential source of soluble, bioavailable phosphorus under the anoxic conditions characteristic of the early Earth and reducing habitable exoplanets lacking continents above sea level (see Supporting Information). Oxygenated basalt weathering experiments were also conducted in order to demonstrate the removal and retention of dissolved $PO_4^{3-}$ as a consequence of the formation of $Fe^{3+}$-oxide minerals upon reaction, providing a comparative analysis of $PO_4^{3-}$ mobility with the anoxic experiments (see Supporting Information).

Reactant basalt was extensively characterized before and after reaction using scanning electron microscopy (SEM), electron microprobe (EMPA) with wavelength-dispersive X-ray spectroscopy (WDS), coupled with synchrotron X-ray fluorescence mapping (SXRF) and Fe K-edge (7.112 keV) X-ray absorption near-edge spectroscopy (XANES) imaging (see Supporting Information). Time-series measurements of solution chemistry from experiments were performed using a Thermo Scientific™ Orion™ PerpHecT™ ROSS™ Combination pH Micro Electrode and Thermo Scientific™ Element™ XR inductively coupled plasma mass spectrometer (ICP-MS). The relative standard deviation (2σ) of the ICP-MS measurements ranges between 6-10% for P, 1-5% for Fe, Mn, and Ni, 1-3% for $^{28}Si$, $^{29}Si$, and $^{30}Si$, and 1-2 % for the major cations, $Ca^{2+}$ and $Mg^{2+}$. The range in temperatures prescribed for the basalt weathering experiments, 15 – 75°C, is similar to temperatures associated with modern low-temperature seafloor springs and is within the range estimated through oxygen isotope composition of carbonate minerals formed within altered oceanic crust. (L. A. Coogan et al., 2019; L.A. Coogan & Gillis, 2013; Gillis & Coogan, 2011; Wheat et al., 2017). The Geochemist's Workbench v. 12.0.4 (Bethke et al., 2018), outfitted with a custom database produced using the DBCreate software package (Kong et al., 2013), was used to calculate the speciation of the time-series solution samples.

## 3. Results and Discussion

Mineralogy and element distributions in reactant basalt are shown in Fig. 1a and Fig. S3-S4, while detailed time-series solution and citrate-bicarbonate-dithionate (CBD) extraction chemistry are given in Fig. 2, Tables S1-S4 and Fig. S6-S11. Analysis by EMPA demonstrates that primary $PO_4^{3-}$ occurs principally as a trace phase within silicates (rather than a more concentrated P-bearing mineral such as apatite), present between 0.01-0.4 wt% as $P_2O_5$ (Fig. 1a). Prior to reductive CBD treatment, some P is also concentrated on altered surfaces in association with secondary Fe-oxides and clay mineral phases (Fig. 1a), as confirmed by SXRF and XANES imaging (Fig. S4). Thermodynamic analysis suggests that the soluble P produced under our experimental conditions is present exclusively as $P^{5+}$ (Fig. 1b), consistent with release of trace substituted P in the silicate lattice during basalt dissolution into a protonated $PO_4^{3-}$ pool (predominantly $H_2PO_4^-$). These observations, together with results from the oxygenated experiments, indicate that the lack of Fe-oxide precipitation during basalt weathering under anoxic conditions leads to efficient release of soluble reactive P into percolating fluids during silicate dissolution.

Time-series solution chemistry also provide strong evidence for effective mobilization of bioavailable P during basalt dissolution under anoxic conditions (Fig. 2, S6, an S8). The rapid decrease in the bulk $^{29}Si/^{28}Si$ ratio of the experimental seawater solutions, together with the roughly constant total dissolved $SiO_2$ concentrations, reflects the dissolution of reactant basalt, mixing of isotopically natural $SiO_2$ with $^{29}SiO_2$-enriched synthetic seawater, and precipitation of Si-bearing secondary minerals (Table S1 and S4 and Fig. 2a). Our tracer results, which allow us to estimate the contribution of basalt-derived dissolved Si to the overall $SiO_2$ budget of the system (see Supporting Information), demonstrate a sharp increase in basalt-derived aqueous Si with reaction progress, asymptotically reaching concentrations of ~35 µmol $kg^{-1}$ after ~1300 hours (Fig. 2b). As silicate dissolution progresses, we observe significant mobility of both dissolved $Fe^{2+}$ and $PO_4^{3-}$ with continued reaction progress (Fig. 2c). These results are in stark contrast to the oxygenated basalt weathering experiments, in which dissolved Fe derived from the dissolution of primary silicate minerals remained below the detection limit while P was either removed from solution (despite high initial dissolved P levels) or remained at steady state levels throughout reaction progress as a consequence of adsorption onto pre-existing and incipiently formed $Fe^{3+}$-oxide minerals (see Table S2 and Fig. S7).

We can further evaluate the potential of submarine basalt weathering under anoxic conditions to serve as a source of bioavailable $PO_4^{3-}$ to the deep ocean by using the time-series $^{29}Si/^{28}Si$ data to quantify the total amount of atmospheric $CO_2$ that would be consumed upon alteration of primary basalt by seawater. We assume a $SiO_2$/Alkalinity ratio (Si/Alk) indicative of the composition of tholeiitic basalt, represented here for simplicity as enstatite (e.g., with Si/Alk = 1.0):

$$MgSiO_3 + CO_2 \rightarrow MgCO_3 + SiO_{2(aq)} \ . \qquad (1)$$

By combining the time-series changes in $^{29}Si/^{28}Si$ with the total dissolved $SiO_2$ of the experimental solution, we can use mass balance to quantify the total $CO_2$ consumed throughout the reaction progress ($\Sigma CO_2$). We can then compare $\Sigma CO_2$ with the total $PO_4^{3-}$ released during basalt dissolution ($\Sigma PO_4^{3-}$), which yields the 'mobility ratio' of bioavailable P released per mol of $CO_2$ consumed during anoxic submarine basalt weathering ($\Sigma PO_4^{3-}/\Sigma CO_2$; Fig 3).

The experiment including the CBD reductive dissolution pretreatment provides the most precise estimate of the $\Sigma PO_4^{3-}/\Sigma CO_2$ ratio during submarine basalt weathering under anoxic conditions, yielding a $\Sigma PO_4^{3-}/\Sigma CO_2$ value of 3.99 ± 1.03 µmol mmol$^{-1}$ (Fig. 3). Combining estimated ranges for modern $CO_2$ outgassing fluxes and riverine $PO_4^{3-}$ fluxes of 5-20 TmolC y$^{-1}$ (L.A. Coogan & Gillis, 2020; Isson et al., 2018; Wallmann & Aloisi, 2012) and 0.032-0.058 TmolP y$^{-1}$ (Ruttenberg, 2014), respectively, and stochastically resampling $10^6$ times from uniform priors yields a median estimated mobility ratio for the modern Earth system of 3.6 µmol mmol$^{-1}$ with a 95% credible interval of 2.26-6.95 µmol mmol$^{-1}$ (Fig. 4a). Our experimental results thus indicate that the $\Sigma PO_4^{3-}/\Sigma CO_2$ value characteristic of submarine basalt weathering under anoxic conditions is of similar magnitude to that estimated for modern weathering of the continental crust (Fig. 4a). This suggests that submarine basalt weathering under anoxic conditions should in many cases be roughly similar in its effectiveness at exporting bioavailable P during $CO_2$ consumption as the weathering of continental crust above sea level.

The $\Sigma PO_4^{3-}/\Sigma CO_2$ value obtained by our experiments can be used to illustrate the potential importance of submarine basalt weathering under anoxic conditions for a hypothetical water-rich silicate planet on which basalt weathering is the primary $CO_2$ sink (Abbot et al., 2012; Kite & Ford, 2018). We envision this scenario as being applicable to portions of Earth's earliest history and to Earth-like volatile-rich exoplanet 'waterworlds' on which oxygenic photosynthesis has evolved. Conceptually, $PO_4^{3-}$ is released to the ocean as volcanic/metamorphic $CO_2$ is consumed during submarine basalt weathering, and some fraction of this bioavailable P passes through the biosphere while the remainder is scavenged. This in turn leads to organic C burial and a release of $O_2$ to the ocean-atmosphere system. The rate of $O_2$ release associated with this process ($J_{O_2}$) is given by:

$$J_{O_2} = J_{volc} \cdot f_{weath} \cdot [\Sigma PO_4/\Sigma CO_2] \cdot (1 - \mathcal{E}_P) \cdot r_{CP} \quad (2)$$

Where $J_{volc}$ represents the rate of volcanic $CO_2$ outgassing, $f_{weath}$ represents the fraction of $CO_2$ removal that is balanced by seafloor basalt alteration, $\varepsilon_P$ represents a global scavenging efficiency, $r_{CP}$ gives the global net C/P ratio for material buried from the oceans, and $\Sigma PO_4/\Sigma CO_2$ is the mobility ratio as described above. Because the values of many of these parameters are uncertain even for the Earth system, these estimates are meant only to provide a conceptual framework that allows us to broadly illustrate the potential large-scale impacts of anoxic submarine P release.

Our results indicate that fluxes of $PO_4^{3-}$ from oceanic crust weathering can potentially support significant biospheric $O_2$ release rates (Fig. 4b). For example, at a volcanic $CO_2$ flux of 20 Tmol $y^{-1}$, a global C/P burial ratio of 300, and a P scavenging efficiency of 50% our mobility ratio results indicate a biospheric $O_2$ flux of ~10 Tmol $y^{-1}$. For comparison, the total net biospheric $O_2$ flux on the modern Earth is on the order of ~10-20 Tmol $O_2$ $y^{-1}$ (Catling & Kasting, 2017), while biospheric $O_2$ fluxes during the Proterozoic following the initial oxygenation of Earth's atmosphere have been estimated to be roughly 2-5 Tmol $O_2$ $y^{-1}$ (Ozaki et al., 2019a). Although the precise values of P scavenging efficiency and global burial C/P ratio are not fully known through Earth's history or across a wide range of planetary scenarios, they are likely to operate inversely to one another (e.g., Reinhard at al., 2017b), such that our simple estimate, though non-unique, is likely to be broadly representative.

## 4. Conclusions

Studies of Earth system evolution and conceptual models used to forecast the emergence and maintenance of biosignatures on volatile-rich exoplanets have neglected the differences in hydrothermal P cycling between oxic and anoxic systems, and have instead implicitly invoked conditions under which P is effectively scavenged through adsorption onto $Fe^{3+}$-oxide minerals formed from the hydration and oxidation of primary silicates in submarine basalt (Korenaga et al., 2017; Unterborn et al., 2018). Our results stand in strong contrast to this prevailing conceptual model, and thus provide impetus to revisit mechanistic models for Earth's early oxygen cycle (e.g. Mills et al. (2014)) and the factors regulating the oxygen cycles of volatile-rich silicate planets more generally. In particular, it will be important for future work to establish the ocean-atmosphere $O_2$ 'threshold' above which oxygenation of the deep oceans attenuates bioavailable P fluxes by initiating widespread $Fe^{2+}$ oxidation within the ocean interior, and the dynamics of P scavenging and global C/P/$O_2$ stoichiometry across a wider range of planetary boundary conditions. Nevertheless, our results clearly suggest that biospheres sustained entirely by bioavailable P released during submarine basalt weathering under anoxic conditions, including that of the earliest Earth, are potentially capable of generating extremely high biogenic gas fluxes that rival or exceed even those of the modern Earth. Anoxic submarine weathering should thus be considered an important component of the large-scale redox balance of terrestrial planets.

**Acknowledgments**

DDS was primarily funded through the Flint Postdoctoral Fellowship provided by Yale University. CTR and NJP acknowledge support from the NASA Interdisciplinary Consortia for Astrobiology Research (ICAR) program. CTR acknowledges support from the NASA Nexus for Exoplanet System Science (NExSS). We also acknowledge the Australian Synchrotron (AS) for awarding beam-time to DDS, BE, JB (AS-Proposal #13283). All datasets for this research are included in this paper (and its supplementary information). Code for the geochemical speciation and P-C-$O_2$ mass balance models will be publicly available on GitHub upon publication.

**FIGURES:**

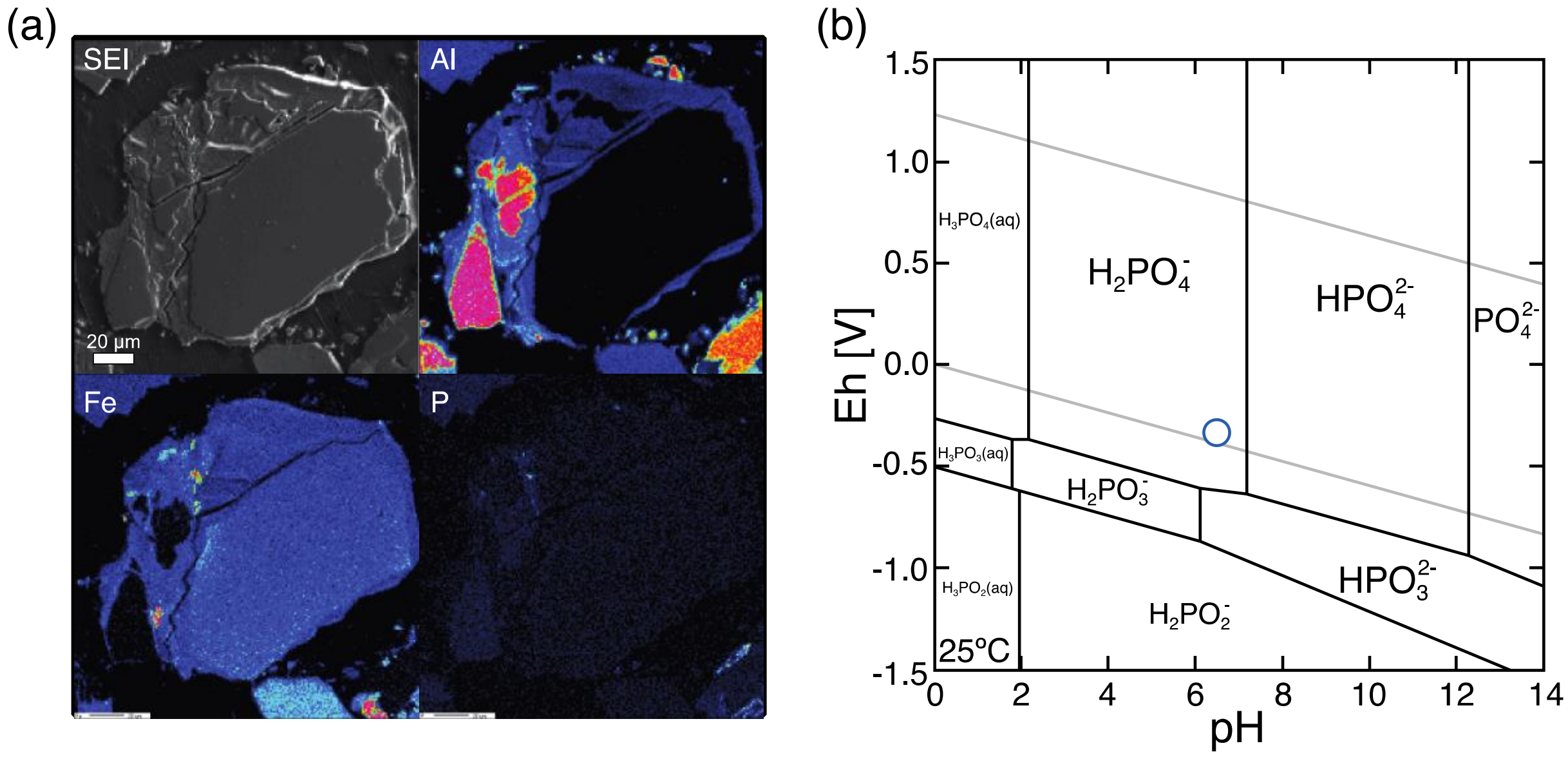

**Figure 1**. (**a**) Representative electron microscopy maps of Al, Fe, and P from grains derived from ground basalt sampled from the Juan de Fuca Ridge used in the seafloor weathering experiments. Phosphorus is concentrated along areas of altered grains that are concentrated in Fe, attributable to a mixture of secondary $Fe^{3+}$-oxide minerals and clays formed upon interaction with seawater at the sampling location. (**b**) Activity diagram showing the speciation of phosphorus in solution as a function of pH and solution redox state (Eh) at 25°C. The open circle shows the conditions of our seafloor weathering experiments, while the grey lines show the stability limits of water. Hot colors towards red represent high concentrations of the element of interest.

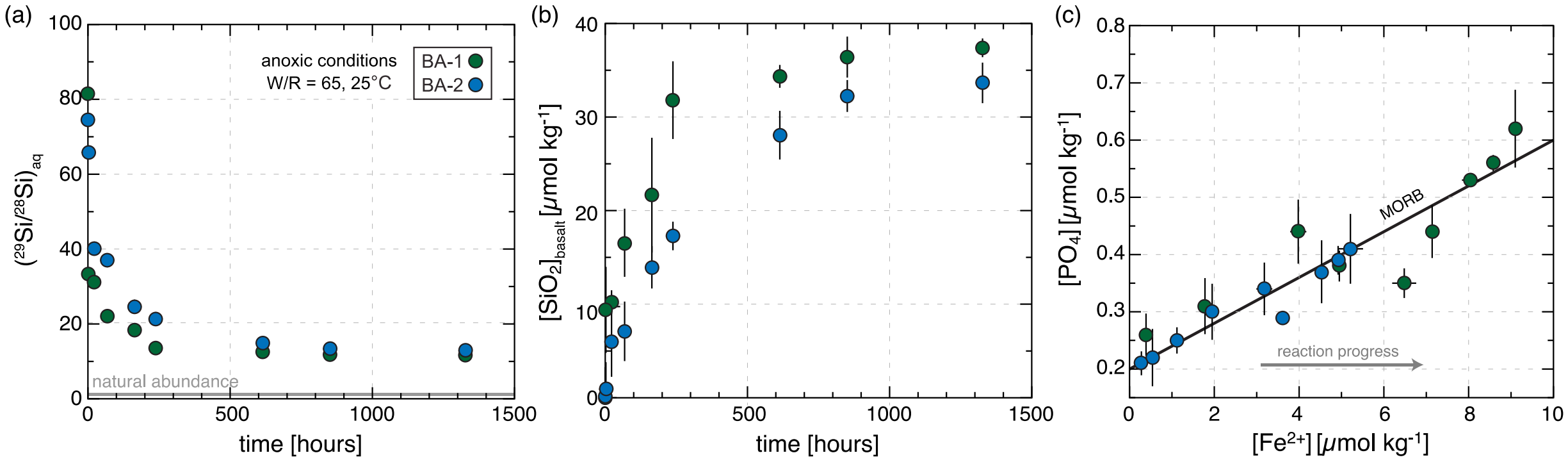

**Figure 2**. Time-series solution chemistry for the anoxic seafloor weathering experiments (BA-1 and BA-2). (**a**) Dissolved $^{29}Si/^{28}Si$ ratio of experimental solutions, showing a rapid drop accompanying the dilution of $^{29}SiO_2$-enriched synthetic seawater by dissolved Si derived from basalt dissolution with a natural abundance $^{29}Si/^{28}Si$ ratio. (**b**) Calculated dissolved $SiO_2$ derived from dissolution of basalt. (**c**) Dissolved $Fe^{2+}$ and $PO_4^{3-}$ concentrations in experimental solutions, showing progressive increase in both species with continued reaction progress. Black line in (**c**) shows the P/Fe ratio expected for dissolution of fresh mid-ocean ridge basalt (MORB).

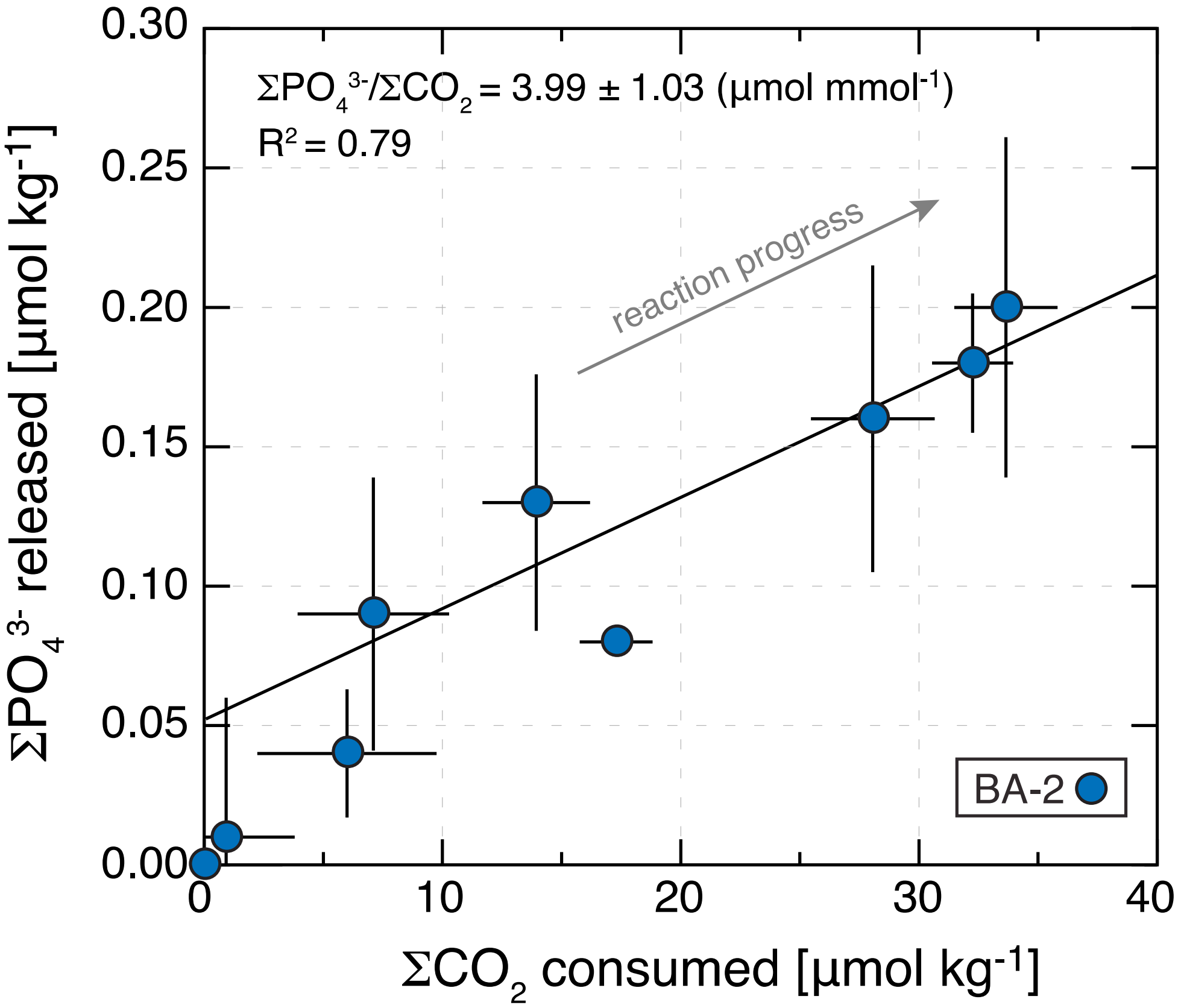

**Figure 3**. Relationship between the amount of $PO_4^{3-}$ released into reactant seawater and the predicted $CO_2$ consumed in experiment BA-2 (including reductive dissolution pre-treatment) upon reaction with basalt under anoxic conditions determined from the time-series $^{29}Si/^{28}Si$ data. Solid line shows linear least-squares regression of the time-series data, the slope of which yields our estimated mobility ratio ($\Sigma PO_4^{3-}/\Sigma CO_2$) for submarine basalt weathering under anoxic conditions, 3.99 ± 1.03 (μmol $mmol^{-1}$) (±1σ).

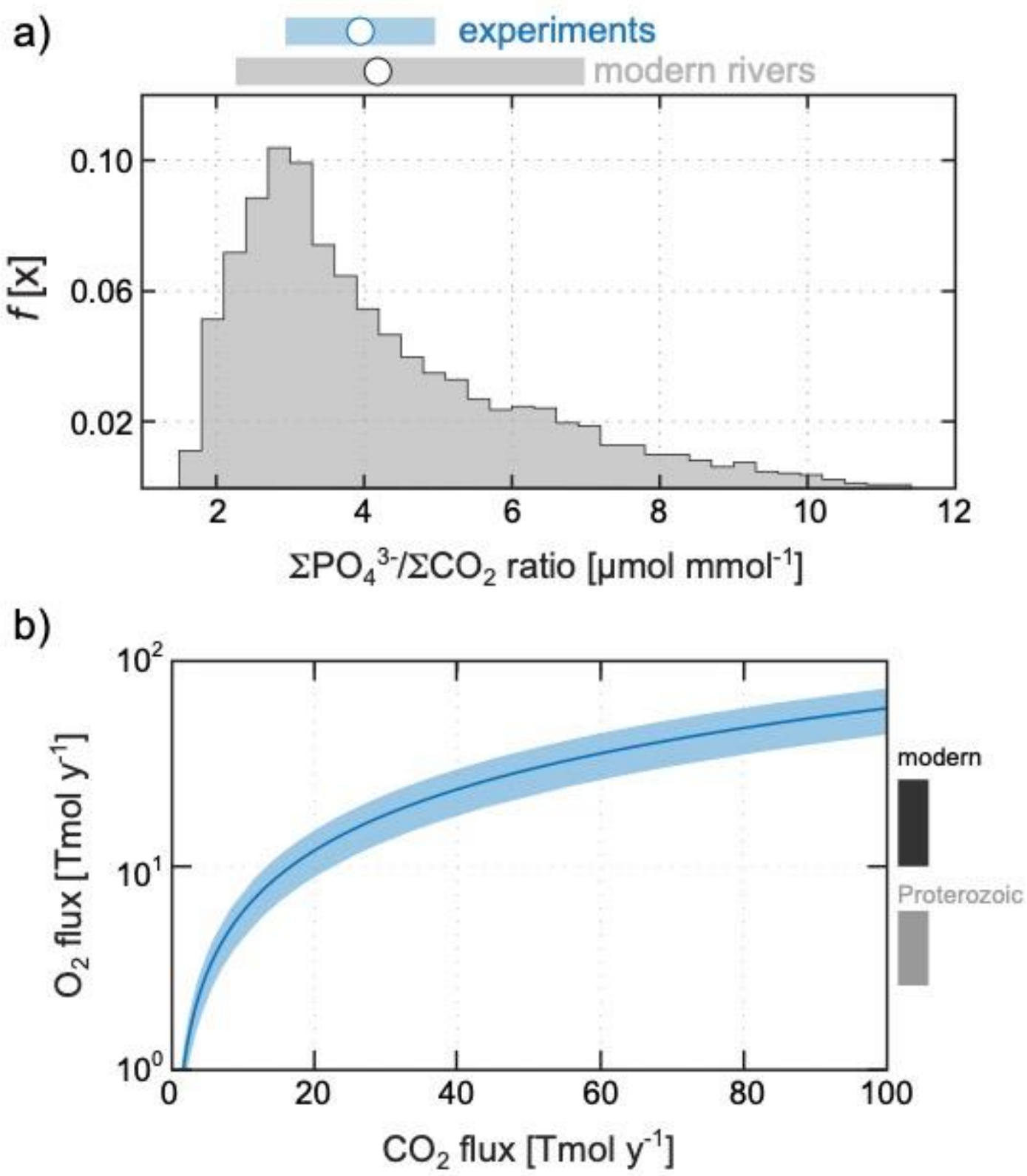

**Figure 4.** Global biogeochemical impacts of submarine basalt weathering under anoxic conditions. (**a**) Comparison of mobility ratios ($\sum PO_4^{3-}/\sum CO_2$) between our experimental results and the modern Earth system, which is dominated by terrestrial weathering of exposed continental landmasses. Grey histogram shows the frequency distribution of resampled estimates for the modern Earth system (see text), with the open circle and shaded bar showing the mean and 95% confidence interval, while the blue circle and shaded bar show the results from our anoxic experiments ($\pm 1\sigma$). (**b**) Estimated biospheric $O_2$ flux as a function of volcanic $CO_2$ input, leading to submarine basalt weathering, P release, and subsequent organic C burial. Calculations assume a P scavenging efficiency of 50% and a global burial C/P ratio of 300 (see text). Blue line and shaded envelope are for the range of mobility ratios shown in (**a**). Ranges for modern and Proterozoic biospheric $O_2$ fluxes are given at right (Catling & Kasting, 2017; Ozaki et al., 2019b).